\documentclass[conference]{IEEEtran}

\usepackage{cite}
\usepackage{graphicx}
\usepackage{paralist}
\usepackage{xcolor}

\begin{document}

\makeatletter
\def\ps@IEEEtitlepagestyle{%
\def\@oddfoot{\parbox{\textwidth}{\footnotesize
Author's version of a paper accepted for publication in Proceedings of the 3rd International Conference on Smart Energy Systems and Technologies (SEST). 
\\
\textcopyright{} 2020 IEEE. 
Personal use of this material is permitted.  
Permission from IEEE must be obtained for all other uses, in any current or future media, including reprinting/republishing this material for advertising or promotional purposes, creating new collective works, for resale or redistribution to servers or lists, or reuse of any copyrighted component of this work in other works.\vspace{1.2em}}
}%
}
\makeatother

\bstctlcite{IEEEexample:BSTcontrol}

\title{Graph-based Model of Smart Grid Architectures}

\author{
\IEEEauthorblockN{%
Benedikt Klaer\IEEEauthorrefmark{1},
{\"O}mer Sen\IEEEauthorrefmark{1},
Dennis van der Velde\IEEEauthorrefmark{1},
Immanuel Hacker\IEEEauthorrefmark{1},
Michael Andres\IEEEauthorrefmark{1},
Martin Henze\IEEEauthorrefmark{3}}

\IEEEauthorblockA{%
\IEEEauthorrefmark{1}\textit{Digital Energy, Fraunhofer FIT,} Aachen, Germany\\
Email: \{benedikt.klaer, oemer.sen, dennis.van.der.velde, immanuel.hacker, michael.andres\}@fit.fraunhofer.de}

\IEEEauthorblockA{%
\IEEEauthorrefmark{3}\textit{Cyber Analysis \& Defense, Fraunhofer FKIE,} Wachtberg, Germany\\
Email: martin.henze@fkie.fraunhofer.de}
}

\maketitle

\begin{abstract}
The rising use of information and communication technology in smart grids likewise increases the risk of failures that endanger the security of power supply, e.g., due to errors in the communication configuration, faulty control algorithms, or cyber-attacks.
Co-simulations can be used to investigate such effects, but require precise modeling of the energy, communication, and information domain within an integrated smart grid infrastructure model.
Given the complexity and lack of detailed publicly available communication network models for smart grid scenarios, there is a need for an automated and systematic approach to creating such coupled models.
In this paper, we present an approach to automatically generate smart grid infrastructure models based on an arbitrary electrical distribution grid model using a generic architectural template.
We demonstrate the applicability and unique features of our approach alongside examples concerning network planning, co-simulation setup, and specification of domain-specific intrusion detection systems.
\end{abstract}


\section{Introduction}
\label{sec:introduction}

The increased penetration of electrical power grids with volatile renewable energies, as well as new load situations, e.g., through electro-mobility, confront grid operators with new challenges:
Bottlenecks due to increased power flow or voltage limits must be identified and remedied by timely active grid operation, leading to increased use of information and communication technology (ICT)~\cite{bian2019performance,velde_medit_2020,Strunz_CIGREBenchmark_2014}.
Failures of this ICT infrastructure seriously endanger the security of power supply, e.g., through the incorrect configuration of communication links, defective algorithms for grid automation, or cyber-attacks~\cite{velde_medit_2020,lee_tlp_2016}.
Consequently, there is a need to
\begin{inparaenum}[(i)]
	\item design and configure ICT infrastructure in power grids to realize topologies that are flexible and resilient to failures,
	\item study the impact of ICT failures and misconfiguration as well as defective algorithms, and
	\item develop resilience and cyber-security approaches that consider both, the energy and ICT domain.
\end{inparaenum}

\noindent\textbf{Motivation and Background.} %
Co-simulations of combined energy and ICT domain are an indispensable tool to analyze these risks and countermeasures at an early stage and enable planning of more resilient power grids~\cite{steinbrink_mosaik_2019}.
To enable such co-simulation of large-scale smart grid scenarios, the consistent and integrated modeling of data points, parameters, interfaces, communication links, as well as ICT and Operational Technology (OT) devices is of utmost importance~\cite{velde_medit_2020,steinbrink_mosaik_2019}.
Focusing on the electrical side of power grids, power grid models are an indispensable tool for grid operators and researchers to plan grid expansion and connection, as well as to solve operational problems~\cite{thurner_pandapower_2018}.
Popular power system simulators and analysis tools such as MATPOWER or pandapower are shipped with different public power grid models~\cite{thurner_pandapower_2018}.
Despite the popularity of modeling approaches for the electrical side of power grids, corresponding models for the ICT side are missing, mainly due to security concerns of grid operators~\cite{velde_medit_2020}.
Thus, synthetic and flexibly constructed ICT infrastructure models based on power grid models provide a valuable alternative.

\noindent\textbf{Relevant Literature.} %
A fundamental foundation for the modeling of smart grids is the Smart Grid Architecture Model (SGAM)~\cite{uslar_applying_2019}.
SGAM allows to describe smart grid system architectures and use cases, with the aim to reveal gaps in smart grid standardization~\cite{Gottschalk2017}.
In its core, SGAM provides an approach to deconstruct the smart grid system landscape into the three dimensions
\begin{inparaenum}[(i)]
\item domains (energy conversion chain),
\item zones, and
\item the interoperability layer~\cite{castro2019modelling, Group2012}.
\end{inparaenum}

The wide applicability of SGAM, enabling distributed simulations and model-driven approaches~\cite{uslar_applying_2019}, is utilized in previous works demonstrating model-based rapid prototyping of smart grids~\cite{andren_engineering_2017}.
Further works are presenting approaches to reduce complexity and improve traceability in the model-driven smart grid application development~\cite{fischinger2019towards}, introducing SGAM-based modeling language for also integrating security by design concept~\cite{neureiter2016domain}.
Furthermore, substation design can be specified by utilizing the common information model together with the IEC 61850 standard through ontology matching~\cite{schumilin2018consistent} within the Query/View/Transformation standard~\cite{lee2017harmonizing}.

However, these studies lack a holistic approach to model smart grid infrastructures, incorporating the technical interoperability specification and configuration of the various components.
W.r.t.\ the practical applicability, a detailed specification of data points, logical component behavior, and ICT network and power grid configuration intertwined in a common model structure is needed.

\noindent\textbf{Contributions and Organization.} %
In this paper, we bridge the gap between existing power grid models and the need to also provide models for the corresponding ICT infrastructure and its network configuration.
To this end, we present an approach to model a graph-based energy information network based on power grid models and smart grid use cases, drawing from the concept of model-driven software engineering~\cite{andren_engineering_2017,fischinger2019towards,Andren_SemanticDrivenDesingMethod_2013}.
The underlying graph structure of our approach enables an automated construction of a smart grid infrastructure model with interoperable network configuration.
Furthermore, the automated creation of data points in OT devices allows for an interoperable configuration, e.g., in co-simulation trials, and serves as a basis for actual deployment.
Our contributions are:
\begin{enumerate}
	\item We provide an analysis of current research challenges regarding missing smart grid ICT models (Section~\ref{sec:problem_analysis}).
 	\item We propose and discuss a structured approach for creating a graph-based infrastructure model consisting of blueprint, modeling, and configuration (Section~\ref{sec:method}).
	\item We show the applicability and benefits of our modeling approach for network planning, co-simulation environments, and ICT security mechanisms (Section~\ref{sec:applicability}).
\end{enumerate}

\section{Problem Analysis} \label{sec:problem_analysis}

With the increasing deployment of IT and OT infrastructure, grid operators, and researchers are facing new challenges w.r.t.\ planning, operation, and security.
In the following, we highlight these challenges using concrete examples and identify requirements to remedy this situation, especially considering integrated modeling of power \emph{and} ICT infrastructure.

\subsection{Planning of ICT Infrastructures} \label{sub:problem_planning}
The transformation of conventionally operated distribution grids to smart grids requires a large-scale expansion of the infrastructure with additional sensor and actuator technology.
The connection and interconnection of new locations to the central SCADA system require forward-looking economic and functional planning of the underlying communication network.
Requirements for communication links result from the various smart grid applications and can be quantified in quality parameters such as bandwidth, latency, jitter, or packet loss rate~\cite{bian2019performance}.
To enable the most economical planning in relation to the usage scenario, mathematical optimization methods can be applied to digital infrastructure models~\cite{Koster2010}.

Thus, planning ICT infrastructures requires knowledge of
\begin{inparaenum}[(i)]
	\item communication topologies and architectures, 
	\item communication link parameter (e.g., bandwidth of single links), as well as
	\item predicted data volume and quality requirements.
\end{inparaenum}

\subsection{Optimization of Operation by Large-scale Co-Simulation} \label{sub:problem_operation}

Simulations are valuable for the development of new methods to optimize the operation of complex energy systems.
Furthermore, negative impacts on a reliable power grid operation due to ICT failures and misconfiguration, as well as cyber-attacks can be safely analyzed through simulations.
Co-simulation enables the execution of multiple simulation models, e.g., a communication network simulator and a power flow solver, in one runtime environment.
Usually, smart grid co-simulation are defined within use cases~\cite{estebsari_SGAM_2019}.

However, current co-simulation approaches are limited by a gap between theoretical use case definition and practical deployment of the simulation environment~\cite{andren_engineering_2017}.
Most importantly, coupling different simulation environments in large-scale scenarios typically requires a prohibitive manual effort.
Furthermore, often, only insufficient data is available either on the energy side or on the communication side to be able to create an environment that sufficiently corresponds to reality.

Thus, to enable the largely automated examination of different methods within co-simulation environments, the following information is required:
\begin{inparaenum}[(i)]
    \item unambiguous definition of the data models required for the individual energy and ICT simulator,
    \item definition of the links between the simulation models, and
    \item automatic generation of distributed device configurations.
\end{inparaenum}

\subsection{Domain-specific Security of Smart Grids} \label{sub:problem_security}

The increased complexity of smart grid ICT networks leads to a multitude of new attack vectors (e.g due to more ICT, new intrusion points, sensitive data exchange, cascade effects) ~\cite{Ponmurugan_IDSSmartGrid_2020}
Past attacks on cyber-physical systems~\cite{langner_stuxnet_2011,lee_tlp_2016,velde_medit_2020} show that semantically correct control and measurement data can still cause critical process states.
Such deceptive manipulation is difficult to detect with conventional security technologies ~\cite{Kim_UndetectableAttack_2013}, thus it is not sufficient to develop countermeasures for energy and ICT separately.

To remedy this situation, firewalls with deep packet inspection could extend security policies on a semantic level and include characteristics of the application layer in case of available consistent information.
Likewise, the availability of specifications of network traffic \emph{and} information on process data would allow us to create process-aware intrusion detection systems.
Finally, for reactive measures, the decision to isolate certain areas or systems requires a risk assessment, which e.g., demands knowledge about operating parameters.

Consequently, to realize security approaches incorporating both, energy and ICT, the following information is needed:
\begin{inparaenum}[(i)]
\item communication relations,
\item protocol information, and
\item valid data points and instruction types for each device.
\end{inparaenum}

\section{Graph-based Modeling Approach} \label{sec:method}
As illustrated by our problem analysis, a wide range of use cases demand integrated infrastructure models of smart grid architectures, also containing information about communication structures, systems, and applications.
We present a structured approach to creating such integrated infrastructure models for smart grids.

To this end, our integrated SGAM-based infrastructure model realizes a contextualized combination of a power grid model with an ICT infrastructure model within a connected graph based on three steps:
\begin{inparaenum}[(i)]
\item blueprint,
\item modeling, and
\item configuration.
\end{inparaenum}
The resulting graph-based model can be used to realize various functionality, e.g., to bootstrap co-simulation environments (cf.\ Section~\ref{sec:applicability}).

\subsection{Blueprint Phase} \label{sub:blueprint_phase}

The first step in creating the infrastructure model is the architectural smart grid design, encompassing the definition of the scenario framework and the considered use cases.
We call this the \emph{blueprint phase}, as it does not refer to a concrete model, but constitutes a general design decision.
The SGAM framework~\cite{Group2012} offers an established standard for the structured description of the scenario framework, providing a wide range of use case definitions~\cite{Gottschalk2017}.

The interdisciplinary relationships of the model specification are reflected in the SGAM interoperability layers.
To illustrate this method, we provide a generic example of the SGAM in Figure~\ref{fig:sgam_example} which breaks down all available information within component, communication, and information layer at the distribution grid level.

\subsubsection{Component Layer} \label{subsub:blueprint_component_layer}
The \emph{component layer} defines assignments between primary technology components in the process zone and ICT components.
Besides metadata for devices, physical connections, and interfaces at the component level are provided.
Connections between the components are represented with a one-to-one cardinality between interfaces.
As the SGAM design is independent of a specific power grid model, the communication network can also only be described conceptually based on SGAM.
In the standard~\cite{Group2012}, the various forms of communication networks are described and assigned to specific technologies depending on where they are used.

\begin{figure} %
    \centerline{\includegraphics[width=\columnwidth]{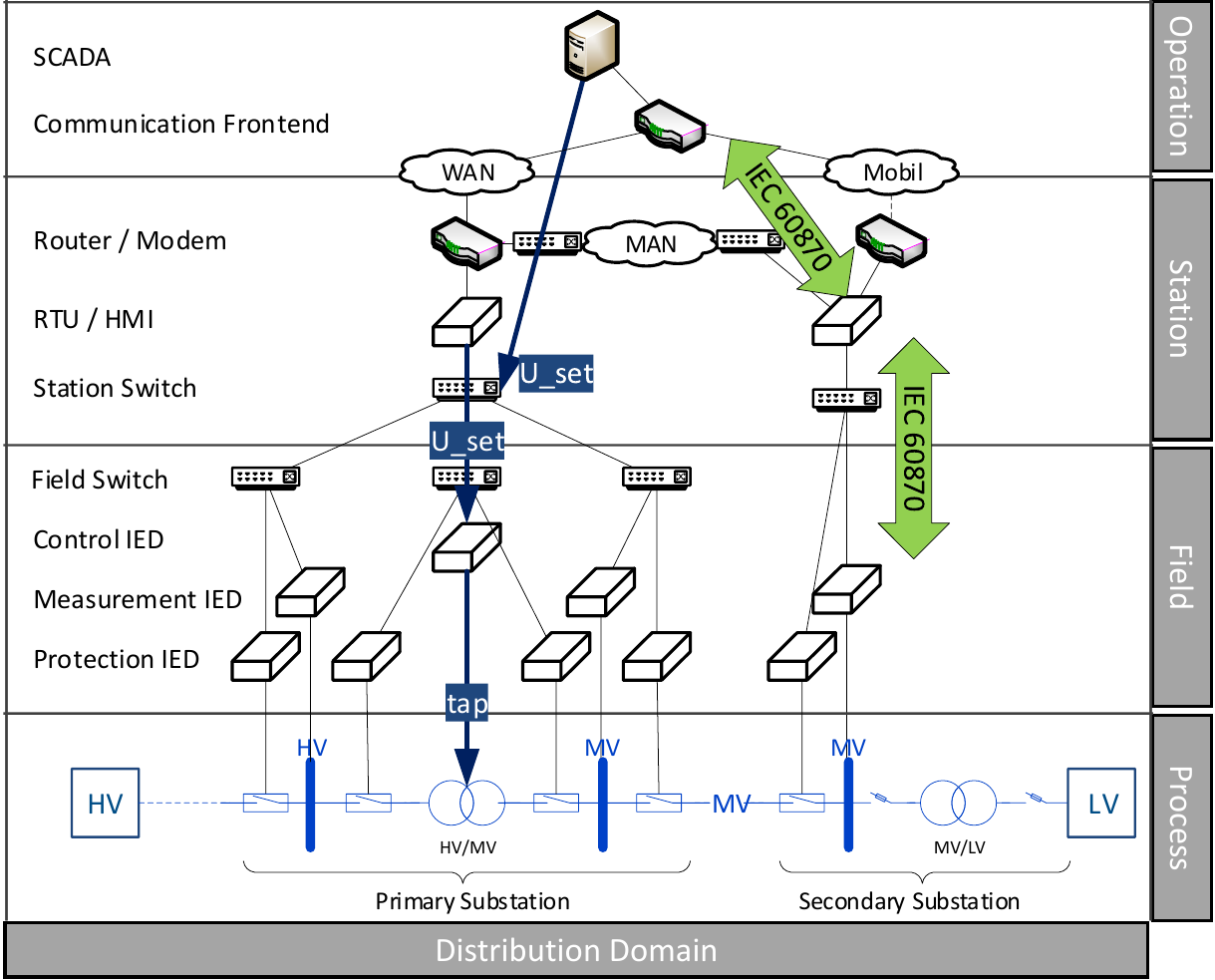}}
    \caption{Generic SGAM example for the blueprint phase, i.e., transformer voltage control use case (communication layer: green, information layer: blue)}
    \label{fig:sgam_example}
    \vspace{-1em}
\end{figure}

The components illustrated in Figure~\ref{fig:sgam_example} are generic and therefore independent of a specific application.
E.g., components at the field-level are subdivided into generic classes of IEDs for control, measurement, and protection.
Specific instantiations of these devices can be found, e.g., in the smart grid standards map~\cite{iec_standardsmap}.
The example shows a generic setup for local area network (LAN) design on substation level with field-based communication links connected by station switches.
Between the station and operation layer, a wide area network (WAN) connects multiple substations with the central SCADA system.
A private fiber-optic or powerline network of the grid operator connects several secondary substations in a metropolitan area network (MAN).
The uplink to the central SCADA system is achieved by a connection to the WAN via the primary substation.
Additionally, connections of secondary substations through mobile communication can be specified.

In summary, the outputs of the component layer are:
\begin{inparaenum}[(i)]
\item ICT components and interfaces,
\item network architecture principle, and
\item network interconnection points.
\end{inparaenum}

\subsubsection{Communication Layer} \label{subsub:blueprint_communication_layer}

The \emph{communication layer} describes the logical connections between smart grid components.
The layer gives information about used protocols and the endpoints for each logical connection.
The example in Figure~\ref{fig:sgam_example} shows communication links between IEDs, RTUs, and SCADA via the IEC 60870 protocol.
Based on the position of a component in the automation hierarchy, we implicitly derive their role as client/master or server/slave concerning other components.
Devices in a higher automation level represent the client/master of a logical connection.
In the context of modeling communication network settings, connections within the communication layer also define required routes and reachability between components.

Hence, the outputs of this layer of the model generation are:
\begin{inparaenum}[(i)]
\item communication protocols,
\item master/slave relationships,
\item network routes, and
\item communication whitelist.
\end{inparaenum}

\subsubsection{Information Layer} \label{subsub:blueprint_information_layer}

Within the \emph{information layer} the data models are specified.
Based on appropriate standards, canonical data models can be defined (e.g., IEC 61968, IEC 61970, IEC 61850).
For data point-oriented protocols (e.g., IEC 60870, Modbus), which are not object-oriented modeled, data points have to be explicitly assigned to typed variables.
In the case of the IEC 60870 protocol, this means an assignment of information object addresses (IOA), cause of transactions (COT), and type identifiers (TypeID) to variables.
Contrary, object-oriented protocols (e.g., IEC 61850) assignments exist implicitly.
In a co-simulation, e.g., the source of the process data is the result of each power flow simulation step.

The exemplary information layer in Figure~\ref{fig:sgam_example} shows a simplified information flow of a transformer voltage control use case.
Within this layer, we define that a primary substation has a transformer tap position as a process data point.
Components on the field-level connected to the transformer of the substation can access its tap position variable.
For voltage control, the SCADA needs to adapt the voltage setpoint on the process level.
This requires an information flow between the SCADA and IED components, which is routed through the substation's RTU.
Therefore, the configuration of these process data points and the assignment to the typed variables are required for all components involved.
Only this way we can determine within the infrastructure model which data points the OT component requires for the use case and whether a valid connection to the primary component (e.g., transformer) exists.

Summarizing, the outputs of the information layer are:
\begin{inparaenum}[(i)]
\item data points,
\item data flow, and
\item source of information.
\end{inparaenum}

\subsubsection{Function and Business Layer}
The \emph{function layer} and the \emph{business layer} of the SGAM model do not provide information to the modeling method and thus do not have to be considered.

\subsection{Modeling Phase} \label{sub:modeling_phase}

\begin{figure}[t]
    \centerline{\includegraphics[width=\columnwidth]{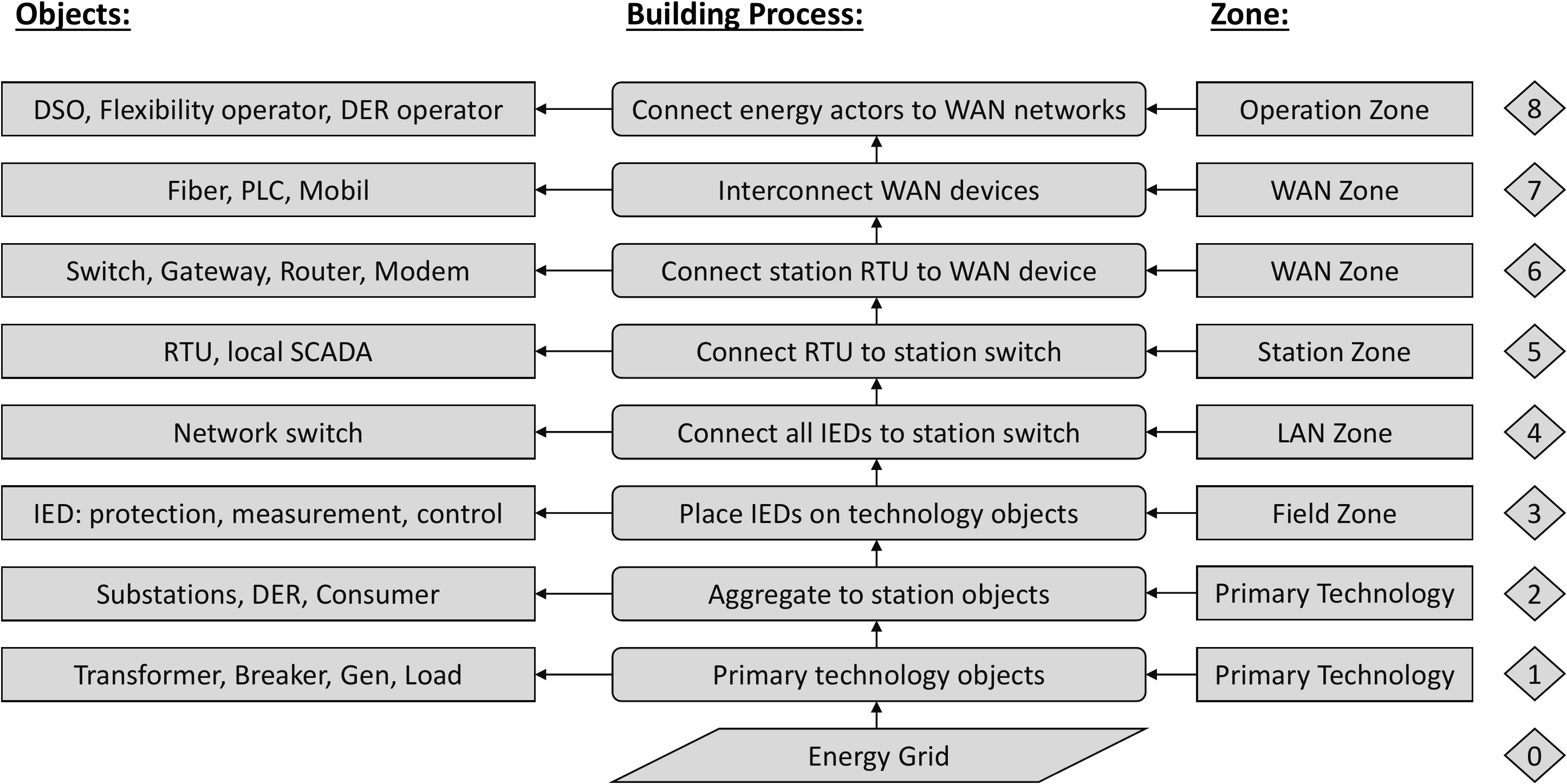}}
    \caption{Flowchart of the bottom-up modeling process for the creation of primary technology and ICT objects as the basis of the digital data model}
    \label{fig:bottom_up}
    \vspace{-1em}
\end{figure}

The modeling phase aims to apply the blueprint to a concrete power grid model to derive a graph representation of an integrated infrastructure model.
A power grid model contains topological information as well as electrical system data and is a graph that represents the lines and systems as edges and nodes.
The application of our blueprint follows a bottom-up approach as illustrated in Figure~\ref{fig:bottom_up}.

\subsubsection{Object oriented Modeling}

In Step 1, we perform object-oriented modeling of the primary technology and instantiate elementary primary objects (e.g., busbars, transformers, loads, and generators) according to the power grid file provided in Step 0.
Subsequently, in Step 2, we perform based on the station specification (component layer of the blueprint) a contextual aggregation of single objects of the primary technology to station objects (e.g., primary/secondary substation, DER, or prosumer).
Afterwards, for each station, a component-specific placement and connection of secondary and protection technology to the primary technology according to the blueprint is performed.
In Step 4, we realize station-wide subnets by placing and connecting local network components to field-level devices.
At the station level, RTU units are placed in Step 5.
The connection of the station level to the WAN (Step 6) follows the specified communication technology (e.g., cellular network, powerline, or fiber).

\subsubsection{WAN Graph Modeling} \label{subsub:wan_graph_modeling}
In Step 7, the WAN is set up as a connecting communication network.
The topology of the communication network can be without explicitly specified deduced from the used communication technologies and the power grid model.
To this end, we identify, independent of concrete technological realization, three communication paradigms:
\begin{inparaenum}[(i)]
    \item unbound to electrical topology (e.g., mobile),
    \item linked to electrical topology (e.g., PLC), or
    \item parallel to electrical topology (e.g., fiber or wire).
\end{inparaenum}

If the WAN infrastructure is \emph{unbound to electrical topology}, e.g., in the case of mobile communication, we first assign mobile radio modems located at the stations to local base stations (Node-B) in cell structures.
The network traffic is then routed via base station controllers within the hierarchical internal network of the communication provider.
Consequently, the grid operator can access the RTUs from a central location, resulting in a star-shaped hierarchical communication infrastructure.

WAN infrastructure \emph{linked to electrical topology}, e.g., powerline communication, uses the existing cable and overhead line network.
Thus, the resulting topology of the communication network is strictly bound to the electrical topology and thus impacted by switching measures of the grid operator.

Finally, WAN infrastructure running \emph{parallel to electrical topology}, e.g., optical fiber cables, is laid parallel to the electrical topology. %
The resulting topology is line-oriented, independent of the switching state or electrical influences, and can be further optimized regarding deployment costs through the usage of minimum spanning tree algorithms (cf.\ Section 2 in~\cite{tin2019applications}).
After the set-up of the WAN in Step 8, the individual actors are connected to their respective networks.

\subsection{Configuration Phase}\label{sub:configuration_phase}

The goal of the configuration phase is to derive the necessary parametrization of the integrated infrastructure model created during the modeling phase (cf.\ Section~\ref{sub:modeling_phase}), i.e., parametrization of the communication network (IP addresses, networking routes) and configuration of process data points.

\begin{figure}[t]
    \centerline{\includegraphics[width=\columnwidth]{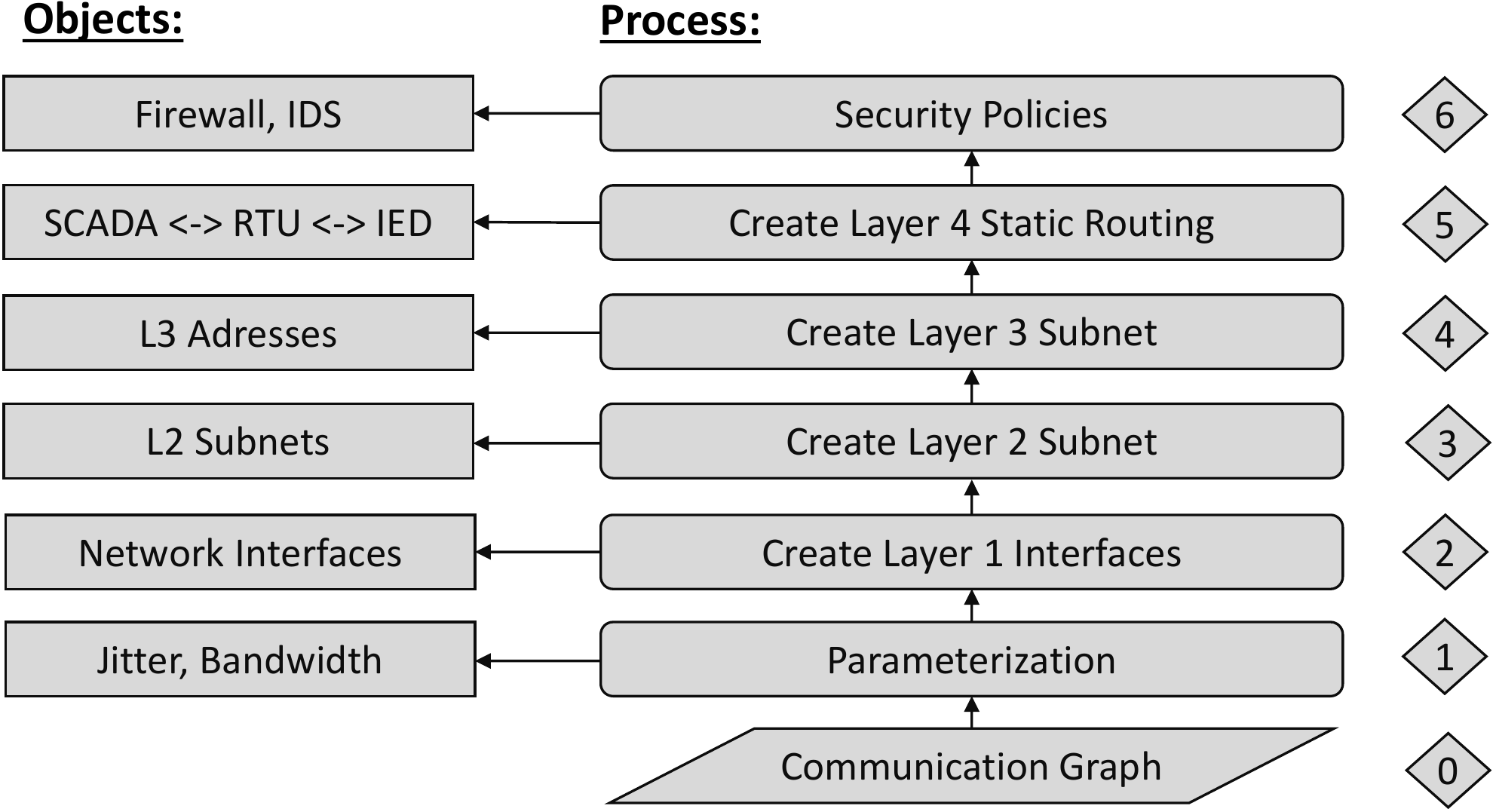}}
    \caption{Process scheme for the consecutive derivation of parameters for the configuration of the communication network}
    \label{fig:wan_config}
    \vspace{-1em}
\end{figure}

\subsubsection{Configuration of Communication Network} \label{subsub:communication_configuration}
The parametrization of the communication network also follows a bottom-up approach as shown in Figure~\ref{fig:wan_config}.
We can directly derive all physical communication links from the graph of the integrated infrastructure model (cf.\ Section~\ref{sub:modeling_phase}).
Initially in Step 1, we parameterize the communication links based in the provided specification regarding quality metrics (e.g., maximum bandwidth, delay, or jitter).
In Step 2, we initialize the communication interfaces on OSI Layer 1 for all components, where the number of interfaces follows the node degree in the graph of the infrastructure model.
During Step 3, we split the graph into OSI Layer 2 subnets at all routing or gateway components and connect the station zone with the operation zone.
Based on these subnets, IP addresses on OSI Layer 3 are allocated to devices in Step 4.
Necessary routes and communication reachabilities are derived from client and server roles defined at the protocol level (cf.\ Section~\ref{sub:blueprint_phase}).
Static routes between components are computed through graph-based shortest path algorithms considering quality metrics (cf.\ Section 3 in~\cite{tin2019applications}).
In Step 5, static routing entries are registered for components with routing functionality on the resulting paths, ensuring connectivity between the subnets.

An additional feature of the infrastructure model is the automated generation of security policies, e.g., in firewalls or intrusion detection systems.
We derive security policies from communication relationships, address assignments, and protocol properties (e.g., used ports), specified by the protocol level in the blueprint phase.
Application layer security can be configured by also providing information about the exchanged process data, collected during data point configuration.

\subsubsection{Data Point Configuration} \label{subsub:datapoint_configuration}

\begin{figure}[t]
    \centerline{\includegraphics[width=\columnwidth]{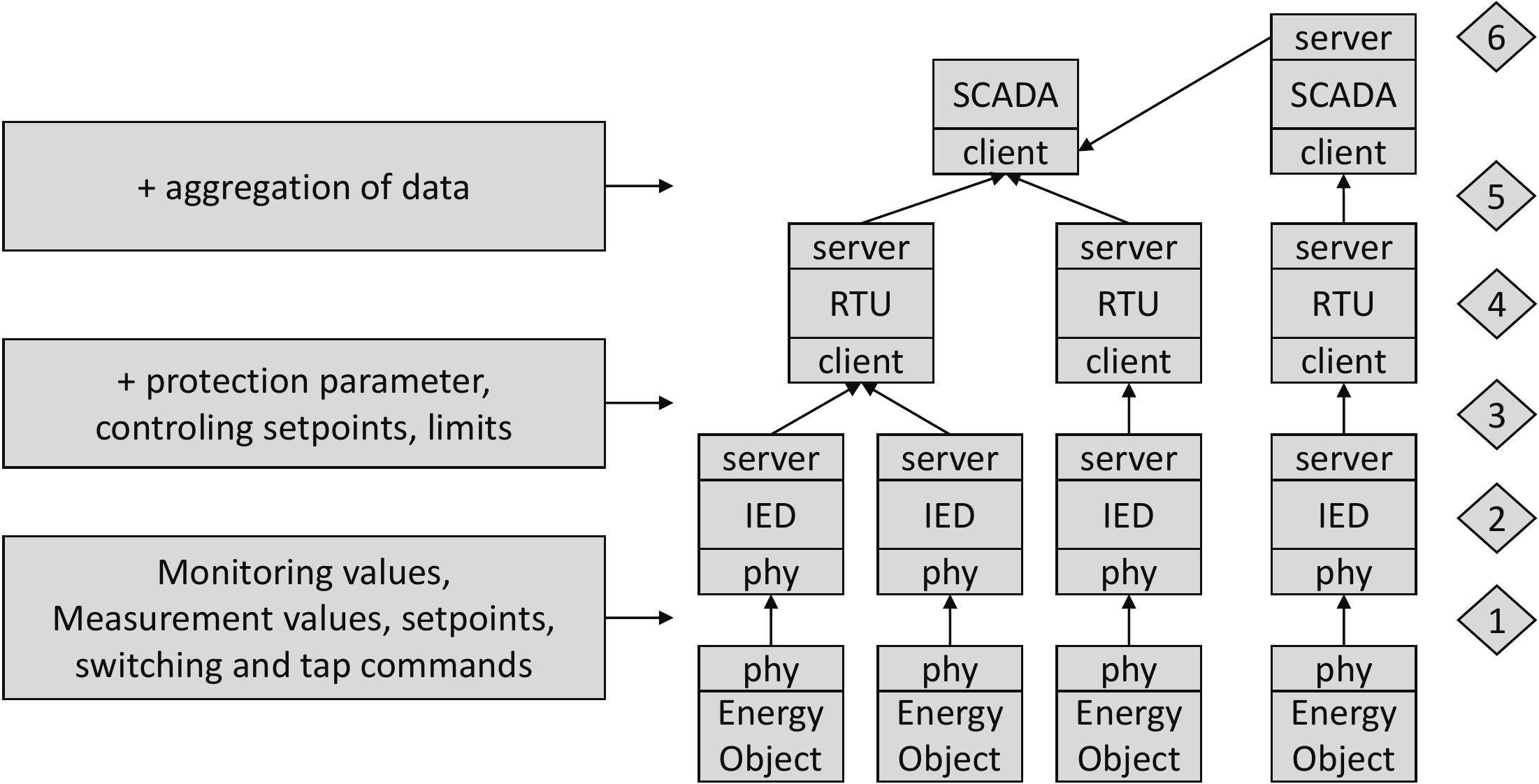}}
    \caption{Process schematic for automatic parametrization of the process data points through hierarchical inheritance between SGAM zone layers}
    \label{fig:hierarchical_parameter}
    \vspace{-1em}
\end{figure}

The results of the blueprint and modeling phase are used for the automated configuration of the process data points, exemplary shown in Figure~\ref{fig:hierarchical_parameter}.

As described in Section~\ref{subsub:blueprint_information_layer}, the information layer defines data of primary technology (e.g., tap position of transformers) and process data points in components (e.g., voltage setpoints).
The availability of process data to field devices (e.g., IED) is defined in Step 1 by the connected physical interface of a primary technology object.
In Step 2, process data points of field devices (e.g., for IEC 60870: IOA, type identifier, COA) are parameterized according to the data model of the information layer.
Using the graph underlying our infrastructure model, we can inherit (Step 3) and aggregate (Step 4) process data points of underlying trees from client applications.

Inheritance continues at the station level and defines the available aggregated process data points for the SCADA system in Step 5.
In Step 6, the inheritance of data points between two SCADA systems is shown and provides utility for flexibility proposals for distribution grid operators.
Further advantages of this structured data point inheritance are:
\begin{inparaenum}[(i)]
\item IEDs receive a correct reference to permitted accesses within the co-simulation,
\item process data points are aligned with energy objects in the grid model (e.g., switches, transformers),
\item data aggregating devices (RTU, SCADA) are automatically configured correctly.
\end{inparaenum}

\subsection{Computational Performance}\label{sub:performance}

As a foundation for a co-simulation case study, we assess the computational performance of our graph-based modeling approach by measuring the time required to automatically create integrated infrastructure models.
We report on the mean processing time and its standard deviation (STD) for the automated build-up process of the infrastructure model over 10 repetitions for 3 medium-voltage grid models.
The first grid model is composed of 4 buses and the corresponding generated infrastructure model consists of 48 nodes, which was computed within 2s (STD 0.4s).
For the second grid model with 12 buses, generating an infrastructure model with 143 nodes takes 8.5s (STD 0.37s), while producing a 318 node large infrastructure model from the CIGRE MV reference grid model~\cite{Strunz_CIGREBenchmark_2014} requires 14.8s (STD 0.8s).
These numbers indicate that our approach to create a contextualized combination of a power grid model with an ICT infrastructure model provides sufficient performance for typical application scenarios.

\section{Applicability} \label{sec:applicability}
Our methodology presented in Section~\ref{sec:method} already shows advantages to the automatable and interoperable construction of a graph-based infrastructure data model.
The applicability of this data model to the results of the problem analysis is explained in the following.

\subsection{Advantages to Network Planning} \label{sub:applicability_planning}

Using the method presented in Section~\ref{sec:method}, communication networks can be constructed based on the power grid model.
Through the explicit data point configuration of each device (cf.\ Section~\ref{subsub:datapoint_configuration}),
as well as the respective communication relationships, the net data volume can be estimated protocol-unspecific for monitoring and control applications depending on the cycle duration, and in the burst case.
Since static routes are part of the specifications, the packet flow is deterministic and the bandwidth of the individual links as well as the time response between hosts can be estimated conservatively over the entire communication graph.
For a more detailed evaluation of the protocol overhead, time behavior, or influence of simultaneities on message queuing, the execution of the scenario in a co-simulation (cf. Section~\ref{sub:applicability_cosim}) is necessary.
Economic feasibility studies for the placement of additional communication lines in case of limit violations can be carried out automatically with the help of graph-based optimization methods (e.g., Steiner tree problem).

\subsection{Deployment of Co-Simulation Environments} \label{sub:applicability_cosim}
\begin{figure}[t]
    \centerline{\includegraphics[width=\columnwidth]{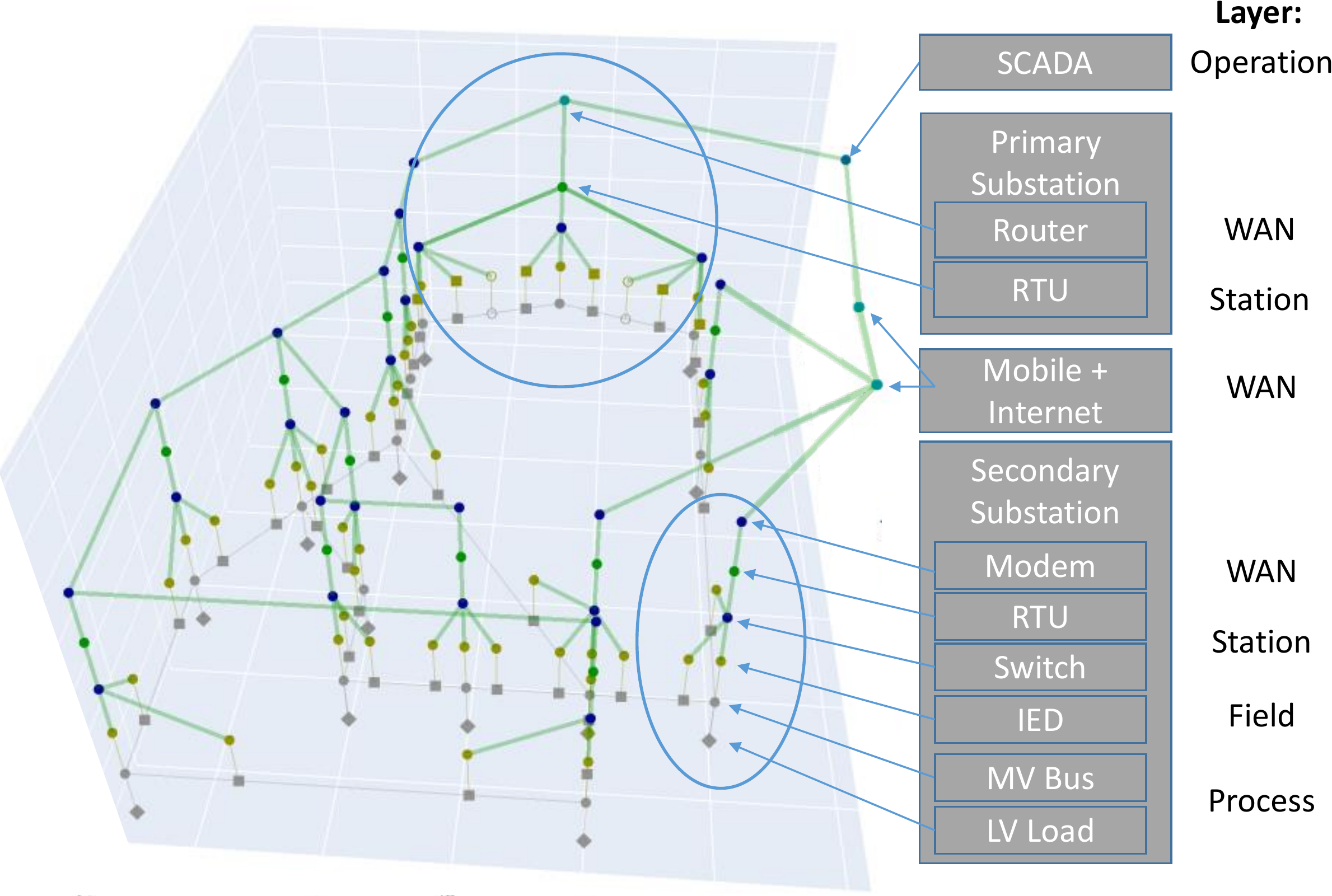}}
    \caption{Visualization of the graph-based data model of the CIGRE MV reference grid based on the template of Figure~\ref{fig:sgam_example}}
    \label{fig:example_grid_graph}
    \vspace{-1em}
\end{figure}
One of the major challenges in realizing co-simulation for operational analysis is the manual effort required to couple different simulation environments, e.g., power flow simulation and communication network simulation, in large-scale scenarios for deployment (cf.\ Section~\ref{sub:problem_operation}).
To showcase the applicability and benefits of our graph-based model, we use it to create a co-simulation of the grid and communication infrastructure of a distribution system operator using pandapower~\cite{thurner_pandapower_2018} as power flow solver, ns-3~\cite{henderson_ns3_2008} as communication network simulator, and mosaik~\cite{steinbrink_mosaik_2019} as co-simulation frameworks.
Starting from a concrete power grid model, we use our approach to derive a graph representation of an integrated infrastructure model.
An exemplary result of our modeling approach based on the publicly available CIGRE MV reference grid model~\cite{Strunz_CIGREBenchmark_2014} is shown in Figure~\ref{fig:example_grid_graph}.
Thus, the resulting graph-based model, presented in Figure~\ref{fig:example_grid_graph}, contains all information necessary to setup the corresponding communication network in ns-3 via a YAML network configuration file and couple power flow and communication network simulation using mosaik.

To this end, all ICT and OT assets (e.g., RTUs or switches) as well as their interconnection can be laid out in ns-3 with the information provided by our model.
The network parametrization in ns-3 (e.g., IP addresses or static routes) then follows from the results of the configuration phase (cf.\ Section~\ref{sub:configuration_phase}).
All information required by mosaik to couple both pandapower and ns-3 is automatically available through the graph-based relationships.
For example, this allows an RTU simulated in ns-3 to read the corresponding measurement values from the electrical grid provided by the pandapower simulator.
Consequently, our graph-based integrated infrastructure modeling allows us to fully automate the creation and deployment of co-simulation environments.

\subsection{Whitelist Configuration for Advanced Security} \label{sub:applicability_security}

Specifications for domain-specific firewalls and intrusion detection can be collected directly through the explicit data model.
Thus, whitelists can be created from the data model and the detection of anomalies and even unknown threads becomes possible.
Possible whitelist configurations through our data model include communication (e.g., link quality, routing, packet flows), authentication (e.g., MAC/IP addresses), and process data (e.g., control, measurement, plausibility).
Different whitelist strategies can be examined automatically and comprehensively against different attacks using a co-simulation environment.

\section{Conclusion} \label{sec:conclusion}
A reliable operation of smart grids requires accompanying scientific studies.
To this end, we first identified challenges in the network planning, co-simulation, and development of further IT security tools for smart grids (Section~\ref{sec:problem_analysis}).
Intending to address these challenges, our integrated infrastructure modeling method allows the construction of complex infrastructure models based on the underlying power grid using a graph-based approach (Section~\ref{sec:method}).
We showcase the applicability and advantages of our approach using concrete references to the identified challenges (Section~\ref{sec:applicability}).
In contrast to existing modeling techniques, we provide a way to consistently and automatically create configurations and parametrization for co-simulations and hardware-in-the-loop testbeds.
Furthermore, dependencies and security concepts between the cyber and energy domain can be directly included in the optimization of smart grid topologies and configuration.
Future work will deal with the automated creation of more secure network structures (e.g., VPN, MPLS) in this framework.
To conclude, our work bridges the gap between existing power grid models and the need to also provide models for the corresponding ICT infrastructure by automatically deriving ICT infrastructure models from given power grid models.

\noindent\textsc{Acknowledgments}\hspace{1em}
This work has partly been funded by the German Federal Ministry for Economic Affairs and Energy (BMWi) under project funding reference 0350028.



\begin{thebibliography}{10}
\providecommand{\url}[1]{#1}
\csname url@samestyle\endcsname
\providecommand{\newblock}{\relax}
\providecommand{\bibinfo}[2]{#2}
\providecommand{\BIBentrySTDinterwordspacing}{\spaceskip=0pt\relax}
\providecommand{\BIBentryALTinterwordstretchfactor}{4}
\providecommand{\BIBentryALTinterwordspacing}{\spaceskip=\fontdimen2\font plus
\BIBentryALTinterwordstretchfactor\fontdimen3\font minus
  \fontdimen4\font\relax}
\providecommand{\BIBforeignlanguage}[2]{{%
\expandafter\ifx\csname l@#1\endcsname\relax
\typeout{** WARNING: IEEEtran.bst: No hyphenation pattern has been}%
\typeout{** loaded for the language `#1'. Using the pattern for}%
\typeout{** the default language instead.}%
\else
\language=\csname l@#1\endcsname
\fi
#2}}
\providecommand{\BIBdecl}{\relax}
\BIBdecl

\bibitem{bian2019performance}
D.~Bian \emph{et~al.}, ``Performance evaluation of communication technologies
  and network structure for smart grid applications,'' \emph{IET
  Communications}, vol.~13, no.~8, 2019.

\bibitem{velde_medit_2020}
D.~van~der Velde \emph{et~al.}, ``{Methods for Actors in the Electric Power
  System to Prevent, Detect and React to ICT Attacks and Failures},'' in
  \emph{IEEE ENERGYCon}, 2020.

\bibitem{Strunz_CIGREBenchmark_2014}
K.~Strunz \emph{et~al.}, \emph{{Benchmark Systems for Network Integration of
  Renewable and Distributed Energy Resources}}.\hskip 1em plus 0.5em minus
  0.4em\relax TF C6.04.02: TB 575, 2014.

\bibitem{lee_tlp_2016}
R.~M. Lee \emph{et~al.}, ``{Analysis of the Cyber Attack on the Ukrainian Power
  Grid},'' E-ISAC, 2016.

\bibitem{steinbrink_mosaik_2019}
C.~Steinbrink \emph{et~al.}, ``{CPES Testing with mosaik: Co-Simulation
  Planning, Execution and Analysis},'' \emph{Applied Sciences}, vol.~9, no.~5,
  2019.

\bibitem{thurner_pandapower_2018}
L.~Thurner \emph{et~al.}, ``{Pandapower---An Open-Source Python Tool for
  Convenient Modeling, Analysis, and Optimization of Electric Power Systems},''
  in \emph{IEEE Trans.\ Power Syst.}, vol.~33, no.~6, 2018.

\bibitem{uslar_applying_2019}
M.~Uslar \emph{et~al.}, ``{Applying the Smart Grid Architecture Model for
  Designing and Validating System-of-Systems in the Power and Energy Domain: A
  European Perspective},'' \emph{Energies}, vol.~12, no.~2, 2019.

\bibitem{Gottschalk2017}
M.~Gottschalk \emph{et~al.}, \emph{The Use Case and Smart Grid Architecture
  Model Approach}.\hskip 1em plus 0.5em minus 0.4em\relax Springer, 2017.

\bibitem{castro2019modelling}
M.~Castro \emph{et~al.}, ``Modelling the transition to distribution system
  operator using the smart grid architecture model,'' 2019.

\bibitem{Group2012}
{CEN-CENELEC-ETSI Smart Grid Coordination Group}, ``{Smart Grid Reference
  Architecture},'' 2012.

\bibitem{andren_engineering_2017}
F.~P. Andr{\'e}n \emph{et~al.}, ``{Engineering Smart Grids: Applying
  Model-Driven Development from Use Case Design to Deployment},''
  \emph{Energies}, vol.~10, no.~3, 2017.

\bibitem{fischinger2019towards}
M.~Fischinger \emph{et~al.}, ``Towards a model-centric approach for developing
  dependable smart grid applications,'' in \emph{ICSRS}, 2019.

\bibitem{neureiter2016domain}
C.~Neureiter \emph{et~al.}, ``Domain specific and model based systems
  engineering in the smart grid as prerequesite for security by design,''
  \emph{Electronics}, vol.~5, no.~2, 2016.

\bibitem{schumilin2018consistent}
A.~Schumilin \emph{et~al.}, ``{A Consistent View of the Smart Grid: Bridging
  the Gap between IEC CIM and IEC 61850},'' in \emph{SEAA}, 2018.

\bibitem{lee2017harmonizing}
B.~Lee and D.-K. Kim, ``Harmonizing iec 61850 and cim for connectivity of
  substation automation,'' \emph{Comput.\ Stand.\ Interfaces}, vol.~50, 2017.

\bibitem{Andren_SemanticDrivenDesingMethod_2013}
F.~Andr{\'e}n \emph{et~al.}, ``{Towards a Semantic Driven Framework for Smart Grid
  Applications: Model-Driven Development using CIM, IEC 61850 and IEC 61499},''
  \emph{Informatik Spektrum}, vol.~36, 2013.

\bibitem{Koster2010}
A.~Koster and X.~Muñoz, \emph{{Graphs and Algorithms in Communication
  Networks}}.\hskip 1em plus 0.5em minus 0.4em\relax Springer, 2010.

\bibitem{estebsari_SGAM_2019}
A.~Estebsari \emph{et~al.}, ``{A SGAM-Based Test Platform to Develop a Scheme
  for Wide Area Measurement-Free Monitoring of Smart Grids under High PV
  Penetration},'' \emph{Energies}, vol.~12, no.~8, 2019.

\bibitem{Ponmurugan_IDSSmartGrid_2020}
P.~Ponmurugan \emph{et~al.}, ``{Intrusion Detection Strategies in Smart
  Grid},'' in \emph{{Design and Analysis of Security Protocol for
  Communication}}, 2020.

\bibitem{langner_stuxnet_2011}
R.~Langner, ``{Stuxnet: Dissecting a Cyberwarfare Weapon},'' \emph{IEEE
  Security \& Privacy}, vol.~9, no.~3, 2011.

\bibitem{Kim_UndetectableAttack_2013}
J.~{Kim} and L.~{Tong}, ``{On Topology Attack of a Smart Grid: Undetectable
  Attacks and Countermeasures},'' \emph{IEEE J.\ Sel.\ Areas Commun.}, vol.~31,
  no.~7, 2013.

\bibitem{iec_standardsmap}
{IEC}, ``{Smart Grid Standards Map},'' \url{http://smartgridstandardsmap.com/}.

\bibitem{tin2019applications}
K.~A. Tin, ``Applications of the shortest spanning tree and path on graph
  theory,'' \emph{Algorithms}, vol.~1, no.~2, pp. 3--6, 2019.

\bibitem{henderson_ns3_2008}
T.~R. Henderson \emph{et~al.}, ``{Network Simulations with the ns-3
  Simulator},'' \emph{ACM SIGCOMM Demonstration}, 2008.

\end{thebibliography}
\end{document}